# Dynamical Control of Interlayer Excitons and Trions in WSe$_2$/Mo$_{0.5}$W$_{0.5}$Se$_2$ Heterobilayer via Tunable Near-Field Cavity

Yeonjeong Koo[1#], Hyeongwoo Lee[1#], Tatiana Ivanova[2], Ali Kefayati[3], Vasili Perebeinos[3], Ekaterina Khestanova[2], Vasily Kravtsov[2*], and Kyoung-Duck Park[1*]

[1]*Department of Physics, Pohang University of Science and Technology (POSTECH), Pohang 37673, Korea*
[2]*School of Physics and Engineering, ITMO University, Saint Petersburg 197101, Russia*
[3]*Department of Electrical Engineering, University at Buffalo, The State University of New York, Buffalo, New York 14260, United States*

Email: *vasily.kravtsov@metalab.ifmo.ru*, *parklab@postech.ac.kr*

Emerging photo-induced excitonic processes in transition metal dichalcogenide (TMD) heterobilayers, e.g., coupling, dephasing, and energy transfer of intra- and inter-layer excitons, allow new opportunities for ultrathin photonic devices. Yet, with the associated large degree of spatial heterogeneity, understanding and controlling their complex competing interactions at the nanoscale remains a challenge. Here, we present an all-round dynamic control of intra- and inter-layer excitonic processes in a WSe$_2$/Mo$_{0.5}$W$_{0.5}$Se$_2$ heterobilayer using multifunctional tip-enhanced photoluminescence (TEPL) spectroscopy. Specifically, we control the radiative recombination path and emission rate, electronic bandgap energy, and neutral to charged exciton conversion with <20 nm spatial resolution in a reversible manner. It is achieved through the tip-induced engineering of Au tip-heterobilayer distance and interlayer distance, GPa scale local pressure, and plasmonic hot-electron injection respectively, with simultaneous spectroscopic TEPL measurements. This unique nano-opto-electro-mechanical control approach provides new strategies for developing versatile nano-excitonic devices based on TMD heterobilayers.

**Introduction**

Stacking atomically thin layers of van der Waals (vdW) materials into bilayer heterostructures provides innovative strategies for the development of next-generation optoelectronic devices [1, 2, 3] and substantially broadens the scope of material physics [4]. A plethora of intriguing phenomena has been already unveiled in vdW bilayers, and they are likely just the tip of the iceberg because many more structures remain unexplored with different chemical composition, stacking sequence and angle, interlayer distance, and other parameters. Hence, considerable efforts are currently focused on uncovering and controlling the inherent physical properties in vdW heterostructures.

In particular, interlayer excitons (IXs), formed by electrons and holes spatially separated in the top and bottom layers of transition metal dichalcogenide (TMD) heterobilayers [5], show a range of distinct properties, which are promising for various optoelectronic applications. The reduced spatial overlap of the electron and hole wavefunctions in IXs brings about reduced radiative decay rates, with corresponding lifetimes up to μs [6], while the interlayer distance and twist angle between the constituent monolayers provide knobs for tuning the IX quantum yield [7]. In addition, the out-of-plane component of the IX dipole moment enables straightforward electric field control. IXs in TMD heterobilayers also provide long-lived valley polarization and coherence [8], circumventing the limits of TMD MLs and enabling practical valleytronic applications. Additionally, the slight lattice mismatch and twist angles in heterobilayers give rise to moiré supperlattices and corresponding confinement potentials that can effectively trap IXs [9]. Therefore, IXs in TMD heterobilayers provide much promise for realizing excitonic integrated circuits [10, 11] and possibly demonstrating high temperature many-body effects, such as Bose-Einstein condensates (BEC) and superfluidity [12].

However, in order to enable practical applications of TMD heterostructures, several major challenges must be overcome, one of which is the large degree of spatial heterogeneity. The underlying processes, e.g., competing interactions of coupling, dephasing, and energy transfer of intra- and inter-layer excitons, arise at the nanoscale and cannot be understood by diffraction-limited optical approaches, calling for near-field optical probing [13, 14, 15, 16]. Furthermore, beyond probing, it is highly important to achieve nanoscale control of local IX properties in TMD heterostructures. Yet dynamic control study of nanoscale properties of IXs with simultaneous nano-spectroscopic measurements has rarely been reported [17, 18].

Here, we demonstrate an all-round dynamic control of intra- and inter-layer excitonic processes in a $WSe_2/Mo_{0.5}W_{0.5}Se_2$ heterobilayer with <20 nm spatial resolution using multifunctional tip-enhanced photoluminescence (TEPL) spectroscopy and imaging. The use of the alloy-based heterobilayer allows us to achieve efficient near-field enhancement for both intra- and inter-layer excitons as their spectral peaks all strongly overlap with the tip-plasmon band while remaining well separated [19]. Through hyperspectral TEPL nano-imaging, we reveal nanoscale inhomogeneities of the IX emission and identify regions of different interlayer coupling strength. At the weak interlayer coupling region, we dynamically control the radiative recombination path and competing emission rates of intra- and inter-layer excitons through the engineering of Au tip-heterobilayer distance and interlayer distance, achieving increase of the IX quantum yield compared to that of Xs. In addition, by applying GPa scale tip-pressure to the heterobilayer, we directly modify its electronic bandstructure, which is demonstrated via IX TEPL energy blueshift and supported by theoretical calculations. Furthermore, through the control of plasmonic hot-electron injection from the Au tip, we convert neutral IXs into charged IX states (trions) in a reversible manner. Our results demonstrate that interlayer excitons in TMD heterobilayers can be accurately controlled in nanoscopic volumes via a near-field approach, which opens up new avenues for the development of compact TMD-based optoelectronic devices and provides insights for studying novel many-body phenomena.

## 1. Experimental configuration for TEPL spectroscopy

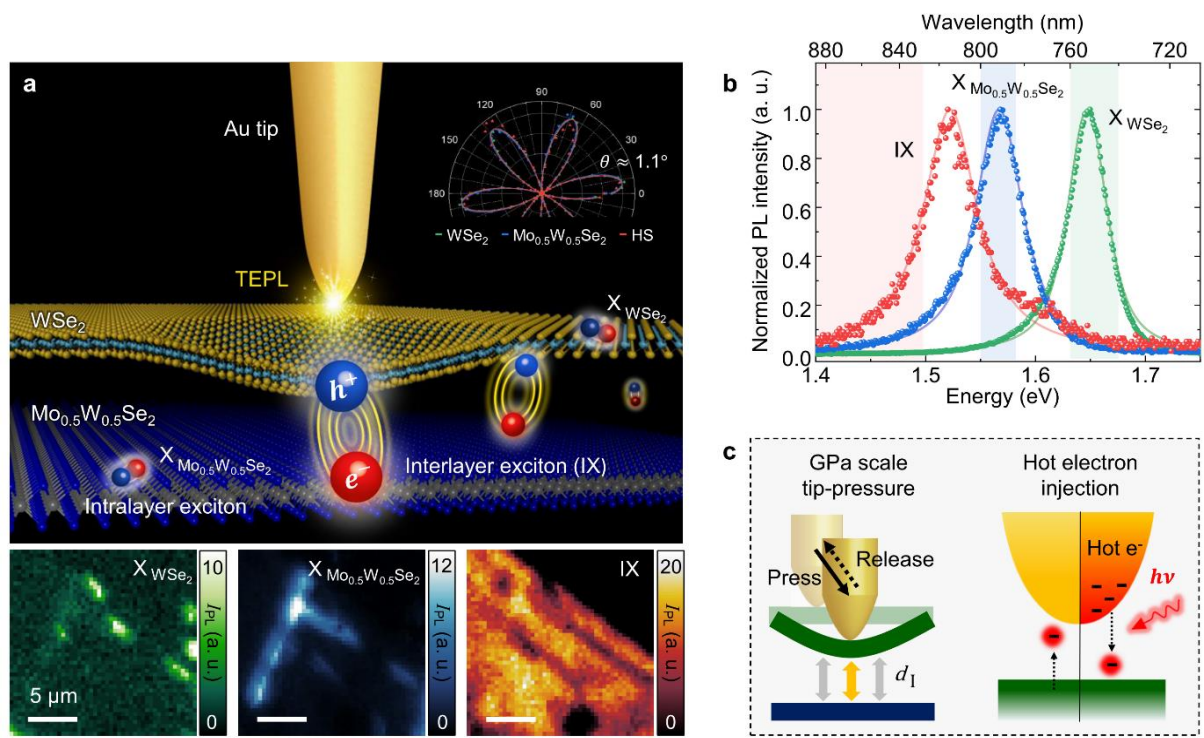

**Fig. 1 Schematic of multifunctional tip-enhanced photoluminescence spectroscopy to dynamically control excitonic processes in TMD heterobilayer.**
**a,** (Top) Illustration of the $WSe_2/Mo_{0.5}W_{0.5}Se_2$ heterobilayer with SHG polarization dependence and the Au tip to probe and control the crystal. (Bottom) Hyperspectral confocal PL images of the heterobilayer for the integrated intensities 740-760 nm (Green, $X_{WSe2}$), 784-800 nm (Blue, $X_{Mo0.5W0.5Se2}$), and 830-900 nm (Red, IX). Scale bars are 5 μm. **b,** Normalized far-field PL spectra of IX, $X_{Mo0.5W0.5Se2}$, and $X_{WSe2}$. Red, blue, and green colored regions are spectral regions used for hyperspectral images in **a** respectively. **c,** Schematic diagram describing the multifunction of TEPL spectroscopy to dynamically control the intralayer- and interlayer-excitonic properties, such as GPa scale tip-pressure and hot electron injection.

We fabricate a WSe$_2$/Mo$_{0.5}$W$_{0.5}$Se$_2$ heterobilayer on a Au film by stacking exfoliated ML flakes with their crystal axes aligned for optimized IX emission. The twist angle is measured via polarization-resolved second-harmonic generation (SHG) spectroscopy to be ~1.1° (see inset of Fig. 1a and Methods). In the assembled heterobilayer, IXs are formed by the spatially separated holes (h$^+$) and electrons (e$^-$) in constituent layers in addition to intralayer excitons X$_{WSe2}$ and X$_{Mo0.5W0.5Se2}$. Hyperspectral far-field PL imaging shows that the spatial distributions of intra- and inter-layer excitons are considerably inhomogeneous at the microscale, as shown in Fig. 1a. Such spatial heterogeneity in van der Waals heterostructures is attributed to the non-uniform interlayer coupling strength, which depends sensitively on local strain-induced deformation and interfacial contamination. Furthermore, on the smaller spatial scales below the diffraction-limit, nanoscale structural deformations such as wrinkles, bubbles, and grain boundaries [20, 21, 22], give rise to complex charge dynamics and interactions with competing recombination processes of intra- and inter-layer excitons.

To develop comprehensive understanding of the nanoscale heterogeneity in the WSe$_2$/Mo$_{0.5}$W$_{0.5}$Se$_2$ heterobilayer and demonstrate its precise control, we develop multifunctional TEPL spectroscopy. We use a radially polarized excitation beam in the bottom-illumination geometry to induce strong out-of-plane optical fields and plasmons at the Au tip-Au film junction (see Methods for more details). The plasmons then couple with the Xs and IXs in the heterobilayer and enhance their PL responses via the Purcell effect [23]. Spectral overlap between the plasmon response and the PL response of intra- and inter-layer excitons in Fig. 1b also give a clear picture of their effective resonant coupling. The tip-sample distance is regulated with a precision of ~0.2 nm using a shear-force feedback loop, with corresponding control on the plasmon enhancement and optical field strength. This allows us to dynamically manipulate the light-matter interactions at the nanoscale with simultaneous spectroscopic TEPL measurements. Fig. 1c shows a schematic of the TEPL spectroscopy experimental configuration, including different multifunctional control modalities, i.e., GPa scale tip-pressure and plasmonic hot carrier injection, as well as tip-induced engineering of the interlayer distance ($d_I$) in a TMD heterobilayer.

## 2. Near-field probing of the nanoscale heterogeneity in a TMD heterobilayer

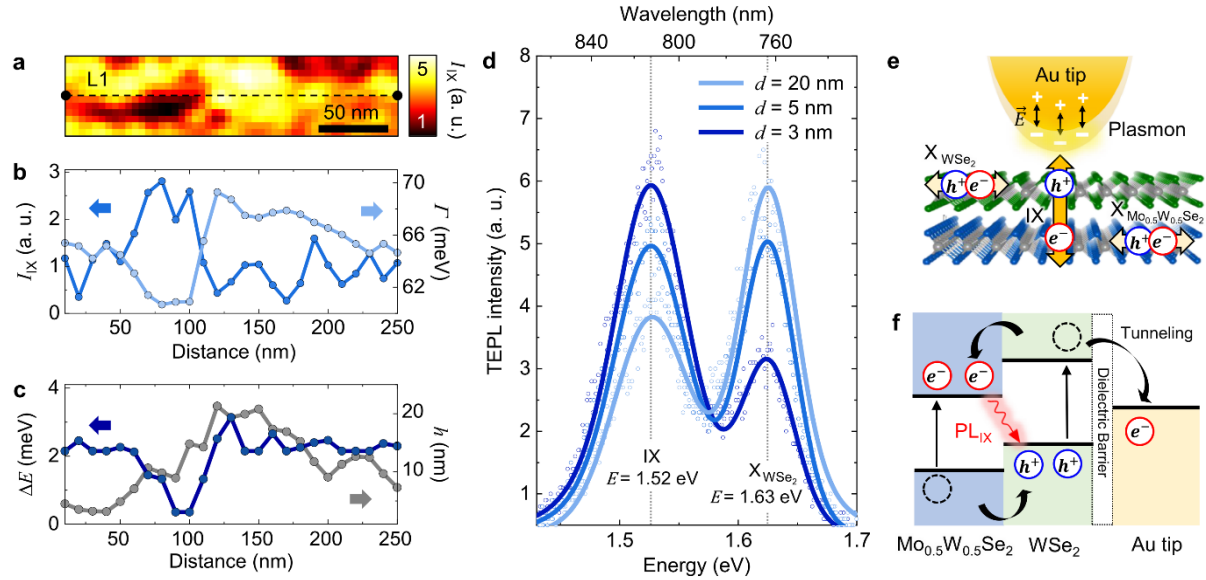

**Fig. 2 Heterogeneous interlayer coupling strength and tip-induced charge transport control.**
**a,** Hyperspectral TEPL image of the heterobilayer exhibiting inhomogeneous IX emission at the nanoscale. **b, c,** Spectroscopic and topographic line profiles for the dashed line L1 in (**a**). Nanoscale spatial heterogeneities in TEPL peak intensity $I_{IX}$, linewidth $\Gamma$, peak energy shift $\Delta E$, and topographic height $h$ are revealed far beyond the diffraction limit. **d,** Evolving TEPL spectra of the heterobilayer as a function of the tip-sample distance $d$. The PL responses of IX (E = 1.52 eV) and $X_{WSe_2}$ (E = 1.63 eV) are acquired with the tip located in the weak interlayer coupling region. **e,** Illustration for the more efficient plasmon-IX (out-of-plane dipole) coupling compared to the plasmon-X (in-plane dipole) coupling when the Au tip closely approaches to the crystal. **f,** Illustration for the type-II band alignment of a $WSe_2/Mo_{0.5}W_{0.5}Se_2$ heterobilayer and the work function of Au tip describing the detailed charge transport mechanisms. This energy transfer mechanism explains our experimental results of increased (decreased) TEPL intensity of interlayer (intralayer) excitons when the tip approaches to the heterobilayer.

To investigate the nanoscale heterogeneity of IXs originated from the non-uniform interlayer coupling strength, we perform hyperspectral TEPL imaging of the heterobilayer, with the experimentally observed spatial distribution of the tip-enhanced IX PL shown in Fig. 2a. In our TEPL scanning, the tip-sample distance $d$ is kept at ~5 nm to minimize tip-induced sample surface modification. To better visualize the spectroscopic information of the inhomogeneous IX distribution and corresponding topography, in Figs. 2b and 2c we present the TEPL intensity $I_{IX}$, peak energy shift $\Delta E$, linewidth $\Gamma$, and height $h$ along the line L1 (indicated in Fig. 2a). The variations in $I_{IX}$ indicate the non-uniform interlayer coupling strength and associated possible changes in the density, oscillator strength, and emission lifetimes of IXs. The regions with higher $I_{IX}$ generally show lower $\Gamma$ and peak energy blueshift. The higher $\Gamma$ in the low-density

IX regions is possibly due to the slight deviation of the IX dipole orientation, since it can cause PL energy variation due to the quantum confinement effect on the interlayer excitonic properties [24]. Similarly, the observed blueshift of the high density IXs is originated from the static electric dipole of IX because the repulsive interactions between the well-oriented IXs cause a mean-field shift, as revealed in previous far-field studies [11]. Note that the height $h$ generally shows an uncorrelated behavior with the spectroscopic line profiles, which means the interlayer coupling strength is not simply characterized by the surface profiling. It should be noted that the whole region of Fig. 2a is measured with ~20 nm spatial resolution by TEPL imaging, which is much smaller than the diffraction-limited beam spot size. Hence, the observed spatio-spectral heterogeneity cannot be investigated using a conventional far-field imaging methods, such as confocal microscopy (See Fig. S1 for the confocal PL image of the same measured area).

We then position the tip in the weak interlayer coupling region and acquire PL spectra of IX and $X_{WSe2}$ as functions of the tip-sample distance $d$, with experimental data for selected distances shown in Fig. 2d. At $d$ = 20 nm, we observe far-field PL spectrum exhibiting IX and $X_{WSe2}$ peaks at $E$= 1.52 eV and 1.63 eV, respectively. The PL peak of $X_{Mo0.5W0.5Se2}$ is not clearly observed due to its low quantum yield. At this relatively large tip-sample distance, the $X_{WSe2}$ peak shows higher PL intensity than the IX peak due to the low interlayer coupling strength. In comparison, at $d$ = 5 nm the intensities of the X and IX PL become similar. Here, the plasmon-exciton coupling in the Au tip-Au film nanocavity is significantly stronger, and, since the plasmonic resonance predominantly enhances out-of-plane optical fields, the PL of the vertically oriented IX dipoles is increased [23]. At the same time, the intralayer excitons, while efficiently excited in the far-field via in-plane polarized fields, show increasingly inhibited PL emission inside the plasmonic cavity at smaller distances $d$. When the tip approaches closely to the heterobilayer with $d$ = 3 nm, the PL intensities of the X and IX peaks are switched, and the IX emission dominates, while its spectral position and shape remain unchanged (See Fig. S2 for more details).

The TEPL enhancement of IXs is attributed to the increased excitation rate and Purcell effect (See SI section 3 for the calculated enhancement factor ~$1.6\times10^3$) [23]. In addition, the tip-induced charge tunneling effect further influences the observed TEPL responses of IXs and Xs [49, 25]. In the near-field regime approaching tip-sample contact, the effective overlap between electron wavefunctions of the Au tip and the heterobilayer can facilitate charge tunneling processes [26] that cause the perturbation of the excitonic system. Fig. 2f illustrates the charge transport mechanism of the type-II band alignment when the tip approaches the

2D crystal surface. Since the Fermi level of Au lies lower than the conduction band minimum energy in WSe$_2$, the electrons at the adjacent WSe$_2$ tunnel into the Au tip. Additionally, the electrons and holes in the heterobilayer are redistributed via interlayer charge transfer. Consequently, the p-doped top layer and the n-doped bottom layer effectively facilitate the IX recombination at the local region with decreasing recombination rate of intralayer excitons, as experimentally confirmed in the result of Fig. 2d.

## 3. Tip-induced nano-engineering of heterobilayer

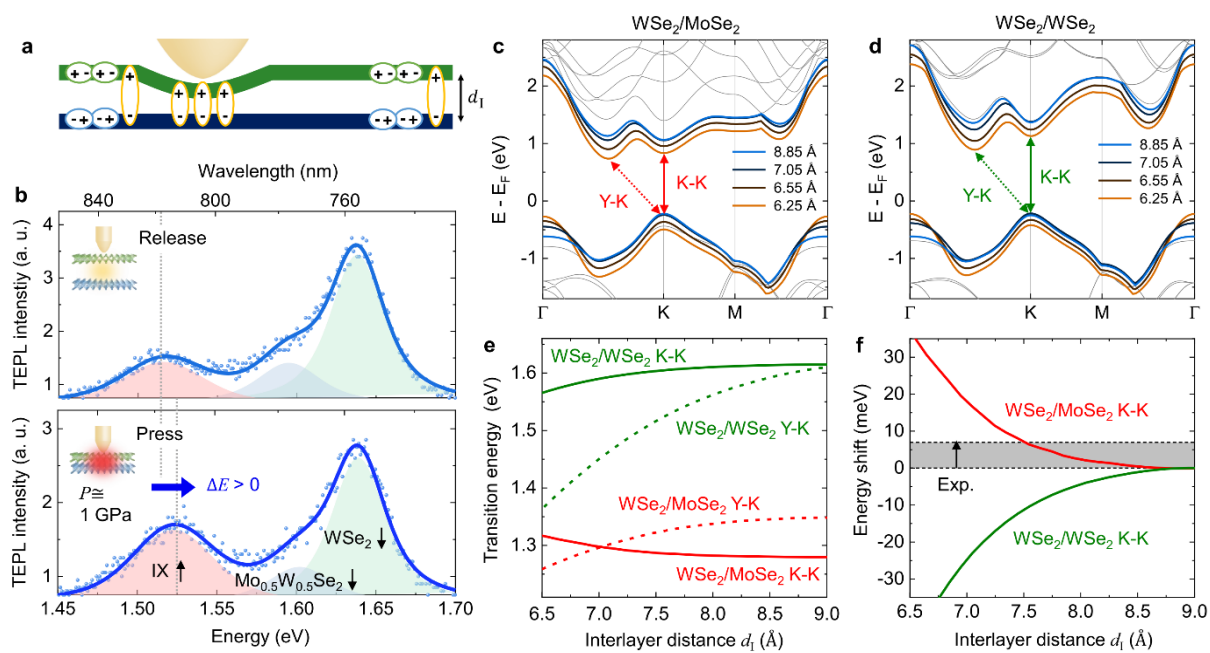

**Fig. 3 Tip-induced control of interlayer coupling strength and electronic bandgap.**
**a**, Schematic illustration of the local tip control of the interlayer distance in a WSe$_2$/Mo$_{0.5}$W$_{0.5}$Se$_2$ heterobilayer. **b**, TEPL spectra before (top) and after (bottom) pressing the heterobilayer with GPa scale tip-pressure, which causes significant modifications in electronic bandstructure. **c**, Electronic bandstructure calculated via DFT for a WSe$_2$/MoSe$_2$ heterobilayer (gray), with the conduction and valence bands for different interlayer distances $d_I$ (from blue for $d_I$ = 8.85 A to brown for $d_I$ = 6.25 A); the direct K-K and indirect Y-K transitions are indicated with arrows. **d**, DFT-calculated bandstructure for a WSe$_2$/WSe$_2$ homobilayer. **e**, Calculated transition energies vs. interlayer distance for the two lowest transitions: direct K-K (solid curves) and indirect Y-K (dashed) in WSe$_2$/MoSe$_2$ (green) and WSe$_2$/WSe$_2$ (red) bilayers. **f**, Calculated energy shifts as functions of interlayer distance for WSe$_2$/MoSe$_2$ (green) and WSe$_2$/WSe$_2$ (red) bilayers; the gray shaded area indicates the experimentally measured tip-induced energy shifts for the interlayer exciton in a WSe$_2$/Mo$_{0.5}$W$_{0.5}$Se$_2$ heterobilayer.

In order to move towards practical opto-electronic device applications of vdW heterobilayers, the nanoscale heterogeneity of IX and X emission should be not only resolved, but also actively controlled. Recently, a few approaches for engineering local exciton properties in 2D heterostructures were demonstrated, for example, via electrostatic field [10, 27] or high magnetic field [28]. Yet, precise nanoscale control of emission beyond the tip-sample distance modulation is a significant challenge [17, 29]. To further extend our tip-induced IX emission control, we present a nano-opto-mechanical tip-pressure engineering approach through the atomic force tip control combined with *in-situ* TEPL spectroscopy. As schematically illustrated in Fig. 3a, the tip exerts local pressure within a ~ 25 nm$^2$ sample area, which is precisely regulated through changing the set-point in a shear-force feedback loop (see Methods). This pressure is expected to cause a local decrease in the interlayer distance and corresponding increase in the interlayer coupling strength. We experimentally verify this behavior by measuring TEPL spectra evolution in a reversible tip-press and -release process. As we demonstrate in Fig. S4 in the Supplementary Information, tip pressure applied to a sample region with initially weak interlayer coupling results in stronger IX emission with simultaneously decreased PL intensities of the peaks corresponding to intralayer excitons in $WSe_2$ and $Mo_{0.5}W_{0.5}Se_2$, which is attributed to the improved interlayer coupling strength [30, 31].

In our previous study, we demonstrated that tip-induced local pressure can exceed 10 GPa owing to its nanoscale tip-sample contact area even though the tip-force is only on the order of 0.1 pN [20, 32]. Here, in the same fashion we induce ~GPa scale tip-pressure in a TMD heterobilayer (see SI section 5 and 6 for the estimation of pressure and compressive strain), which directly modifies its crystal structure and electronic bandstructure, resulting in the modified IX emission properties. Fig. 3b shows the modified TEPL spectra before (top panel) and after (bottom panel) inducing ~GPa scale tip-pressure in the $WSe_2/Mo_{0.5}W_{0.5}Se_2$ heterobilayer. The spectra are decomposed into 3 peaks corresponding to IX, $X_{WSe2}$, and $X_{Mo0.5W0.5Se2}$ via fitting by Lorentzian functions. In addition to the increase in the IX/X PL ratio discussed earlier, the IX TEPL peak exhibits a clearly discernible blueshift of ~7 meV. In order to clarify the physical origin of the observed spectral changes, we simulate the associated electronic bandstructure modification with decreasing interlayer distance using density functional theory (DFT) calculations as described in Methods. In the calculations, we consider two limiting cases of a general alloy-based bilayer $WSe_2/Mo_{1-x}W_xSe_2$ with x = 0 and x = 1, where x represents the relative concentration of W atoms in the alloy layer.

The calculated electronic bandstructures for x = 0 (heterobilayer $WSe_2/MoSe_2$) and x = 1 (homobilayer $WSe_2/WSe_2$) with equilibrium interlayer distance $d_I$ = 8.85 Å, corresponding to unstrained structures, are shown in gray color in Fig. 3c and 3d, respectively. The lower conduction band and upper valence band are highlighted with the blue curves and reveal that the two lowest-energy optical excitations correspond to the momentum-direct K-K (solid arrows) and momentum-indirect Y-K (dashed arrows) transitions. The calculated transformation of the conduction and valence bands with decreasing interlayer distance $d_I$ is shown in Fig. 3c, d with curves changing color from blue ($d_I$ = 8.85 Å) to brown ($d_I$ = 6.25 Å). The corresponding extracted interlayer distance dependencies of the K-K and Y-K transition energies are plotted in Fig. 3e. As observed in the figure, the energies of the lowest optical transitions in $WSe_2/MoSe_2$ and $WSe_2/WSe_2$ exhibit opposite trends with the decreasing interlayer distance, which is highlighted further in Fig. 3f, where the distance-dependent energy shifts are plotted instead of absolute transition energies. While for $WSe_2/MoSe_2$ our DFT calculations predict blueshift of the transition energy, redshifts on a similar scale are predicted for for $WSe_2/WSe_2$. We note that due to the crossover from the direct K-K to indirect Y-K transition in $WSe_2/MoSe_2$ (Fig. 3e, red curves) at $d_I$ ~ 7 Å, the highest predicted blueshifts are limited to 15-20 meV. Considering that for the studied bilayer $WSe_2/Mo_{0.5}W_{0.5}Se_2$ (x = 0.5) the expected energy shifts are smaller as they lie in between those for the x = 0 and x = 1 structures, our experimentally observed energy shift of 7 meV (indicated with a dashed line in Fig. 3f) is at the higher end of the calculated range of values. This can be associated with several factors, including the unknown local stoichiometry of the alloy layer, initial inhomogeneous local strain, and the strain-dependent binding energies [33] of interlayer excitons, which are not accounted for in DFT calculations.

## 4. Tip-induced hot-electron injection control of charged IX

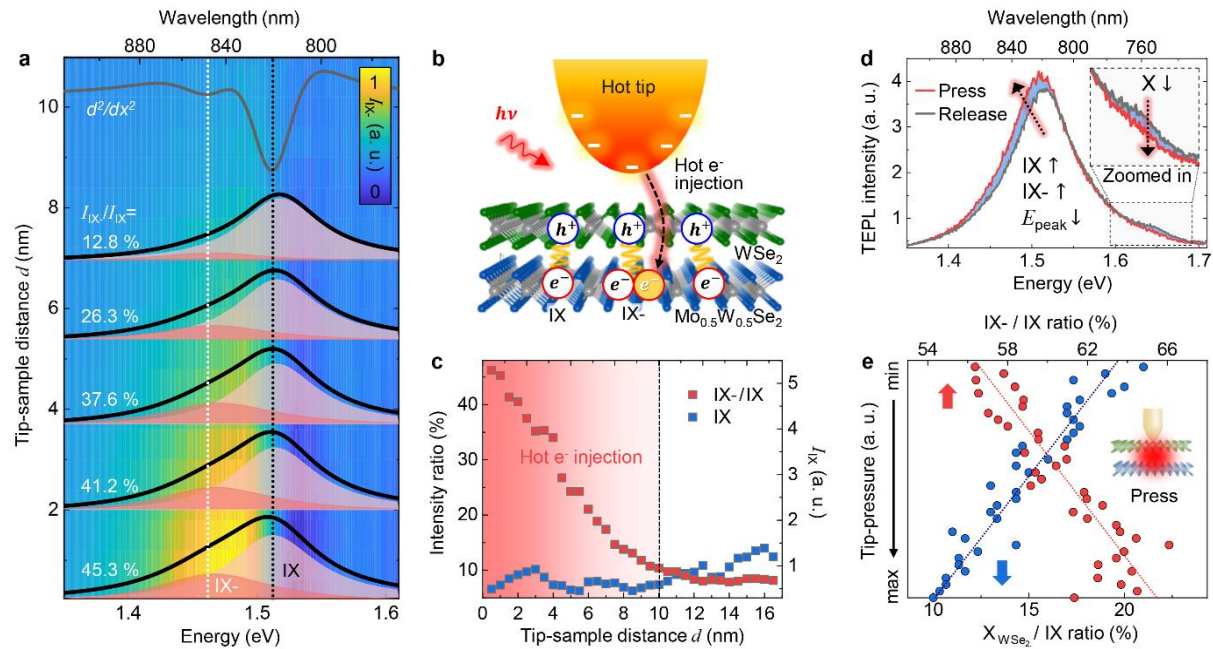

**Fig. 4 Inducing and control of interlayer trions via hot electron injection of the Au tip.**

**a**, TEPL spectra of the heterobilayer (black) as a function of tip-sample distance $d$ at the high excitation power ($\approx 10^9$ W/m$^2$). Interlayer trion (IX-, white dashed line) and interlayer exciton (IX, black dashed line) peaks are assigned by the second derivative of the TEPL spectrum at $d = 1$ nm ($d^2/dx^2$, gray). TEPL intensity ratio of IX- (red) and IX (light red) is derived for the selected spectra via curve fittings with a Voigt function. Subtracted TEPL spectra exhibiting apparent emerging of IX- peak (contour plot). The background contour image exhibits the evolution of the IX- peak obtained by subtracting the TEPL spectrum without hot e- injection (at $d > 10$ nm) from the distance-dependent TEPL spectra. **b**, Illustration for the tip-induced hot-electron injection process, which stimulates the IX- generation in the heterobilayer. **c**, Changes of the IX-/IX ratio (red) and the IX intensity (blue) as a function of tip-sample distance $d$. **d**, TEPL spectra when the Au tip presses (red) and releases (black) the heterobilayer. **e**, Changes of TEPL intensity ratios for IX-/IX (red) and $X_{WSe_2}$/IX (blue) when the Au tip presses the crystal.

The experimental results reported so far have been measured at relatively low values of excitation power ($\approx 10^8$ W/m$^2$). By significantly increasing the excitation power, we can explore a different regime, characterized by electron transport from the Au tip to the heterobilayer, which is due to the hot electron (e-) generation at the plasmonic tip and subsequent injection into the Mo$_{0.5}$W$_{0.5}$Se$_2$ (conduction band of IX) [34, 35, 36]. Our measurements of the excitation power dependent IX PL confirm the increased charged interlayer exciton (IX-) density in contrast to the saturating neutral IX density at the high-power regime attributed to the hot carrier injection [37, 38] (see SI section 7 for more details). By approaching the plasmonic *hot tip* with a strongly localized field close to the heterobilayer, we achieve the dynamic local control of the interlayer trion formation and recombination rate in the near-field regime. To demonstrate such control, we locate the Au tip in the high-quality crystal region with a high interlayer coupling strength exhibiting the dominant IX emission and investigate the tip-induced hot e- injection effect at the high excitation power ($\approx 10^9$ W/m$^2$). Fig. 4a shows the selected TEPL spectra as a function of the tip-sample distance $d$. At $d = 1$ nm, in addition to the neutral IX peak at $E = 1.51$ eV, a pronounced spectral shoulder emerges at the low energy, which is attributed to the charged interlayer exciton or interlayer trion (IX-) peak induced by the hot e- injection. To quantify the IX-/IX PL intensity ratio as a function of $d$, we deconvolute the TEPL spectra into the IX and IX- peaks by the Lorentzian and calculate the integrated intensity of each spectrum. As indicated in Fig. 4a, the IX-/IX ratio increases up to ~45.3 % as the *hot tip* approaches the crystal. The evolution of the IX- peak is also clearly observed in the overlaid false color image, which is obtained by subtracting the TEPL spectrum at a large distance $d > 10$ nm, where the effect of hot electron transfer vanishes, from the distance-dependent TEPL spectra.

The mechanism of the tip-induced IX- generation is schematically illustrated in Fig. 4b. The hot electrons injected into the heterobilayer within the nanoscale region under the plasmonic tip bind with the neutral IXs and form the IX-. Since this process becomes increasingly efficient as the tip approaches the heterobilayer, the local density of the neutral IX is not significantly increased (see Fig. S7 in Supplementary Information for more details). To demonstrate this behavior from the result of Fig. 4a, we plot the integrated IX intensity (blue) and the IX-/IX intensity ratio (red) as a function of distance $d$ in Fig. 4c. The hot $e^-$-induced IX- generation starts to appear at distances $d < 10$ nm. This experimentally observed threshold distance ($d \approx 10$ nm) for the plasmonic hot $e^-$ injection is in good agreement with the results of previous surface-enhanced Raman spectroscopy studies for molecular samples [39]. In addition, the IX-/IX ratio dramatically increases as $d$ decreases while the neutral IX intensity shows no particular change indicating a highly efficient conversion from the neutral IX to IX- under the plasmonic tip as expected.

We further enhance and control the charged IX emission by applying GPa-scale pressure with the tip under high-power excitation. As shown in Fig. 4d, under the tip-induced pressure the contributions to the total TEPL intensity from both IX- and IX peaks are increased, which we attribute to the higher interlayer coupling strength and correspondingly increased recombination rate for both neutral and charged IX species. This is accompanied by redshift of the IX TEPL spectrum and increased linewidth, which is in contrast to the observed blueshift of the TEPL spectrum at low excitation powers presented in Fig. 3b. Additionally, we observe that the TEPL intensity of intralayer excitons ($X_{WSe2}$) is decreased, which is naturally understood from the competing recombination process between the intra- and inter-layer excitons, as discussed earlier with regard to the data presented in Fig. 2. Furthermore, the precise modification of IX, IX- and X emissions is clearly demonstrated in Fig. 4e. When we press the sample with GPa-scale tip pressure, the TEPL intensity ratios for IX-/IX (red) and $X_{WSe2}$/IX (blue) show opposite behaviors with pressure. This result shows a distinct advantage of our work compared to the previous hot $e^-$ injection studies, i.e., the ability to dynamically control the hot $e^-$ density and the corresponding IX- conversion rate. By regulating the tip-sample distance precisely (~0.2 nm [23]) using the scanning probe tip, we can control the hot $e^-$ injection at the nanoscale in a fully reversible manner, which was not possible in the previous studies [40, 41] (See Fig. S8 for demonstration of reversible control).

**Conclusion**

In summary, we have investigated the nanoscale heterogeneity of the interlayer coupling strength in an aligned WSe$_2$/Mo$_{0.5}$W$_{0.5}$Se$_2$ heterobilayer and demonstrated active control of its emission via multifunctional TEPL spectroscopy inside a plasmonic tip-substrate cavity in two distinct power regimes. At low excitation powers, we control the interplay between the intralayer and neutral interlayer exciton PL via distance-tunable Purcell enhancement, where IX emission becomes dominant at small tip-sample distances. At high excitation powers, the plasmonic tip acts as a source of hot electrons, which are injected into the heterobilayer and facilitate formation of interlayer trions with distance-tunable efficiency. Beyond the simple control of interlayer excitons via tip-sample distance modulation, we reversibly modify their spectral response via applying nano-localized tip-induced GPa-scale pressure. We support the observed local pressure-dependent IX spectral evolution with DFT simulations, which provide insights into interlayer distance dependent band structure in aligned TMD bilayers. The presented results demonstrate new approaches to study the nanoscale heterogeneity of the interlayer exciton response in TMD heterobilayers and suggest ways to control that response within nanoscopic sample areas. This manifests an important step towards the development of next-generation optoelectronic devices and investigation of novel many-body effects with TMD-based heterobilayers.

**Acknowledgments**

The reported study was funded by RFBR and National Research Foundation of Korea according to the research project 19-52-51010.

**Methods**

**Sample preparation**

Cover glass (170 um thickness) was ultrasonicated in acetone and isopropanol for 10 mins each and cleaned again by O$_2$ plasma treatment for 10 mins. Then, a Cr adhesion layer (2 nm thickness) and an Au film (9 nm thickness) were deposited subsequently on the glass with a rate of 0.1 Å/s each at the base pressure of ~10$^{-6}$ torr using a conventional thermal evaporator. The prepared substrate was covered with a 0.5 nm thick layer of Al$_2$O$_3$ via atomic layer deposition.

TMD monolayers (Mo$_{0.5}$W$_{0.5}$Se$_2$, WSe$_2$) were mechanically exfoliated from corresponding bulk crystals (HQ Graphene) onto polydimethylsiloxane (PDMS) stamps. For better homogeneity of the target heterobilayer, the monolayers were exposed to UV light [42] for 10 minutes. To achieve accurate layer alignment in the heterobilayer, the directions of crystallographic axes for the monolayers were determined from polarization-resolved second-harmonic generation (SHG) measurements under excitation with laser pulses of 1200 nm center wavelength and 100 fs duration. The WSe$_2$ and Mo$_{0.5}$W$_{0.5}$Se$_2$ monolayers were then stacked together on a PDMS substrate with their crystallographic axes aligned via dry transfer at a temperature of 60 °C. The twist angle between the monolayers in the resulting heterobilayer was measured again with polarization-resolved SHG. Finally, the heterobilayer was placed onto the Au-covered substrate for near-field measurements via dry transfer.

**Multi-functional TEPL spectroscopy and imaging setup**
Multi-functional TEPL spectroscopy is based on the bottom-illumination mode confocal optics setup combined with shear-force AFM using the Au tip. For the excitation beam, He-Ne laser ($\lambda$ = 632.8 nm, optical power $P$ of $\leq 0.5$ mW) was was passed through a radial polarizer and then focused at the Au tip-Au film junction by an oil immersion objective lens (PLN100x, 1.25 NA, Olympus). The radial polarizer was used to make vertically polarized beam component as large as possible at the tip apex which leads to effective coupling of exciton and cavity plasmon inducing highly enhanced TEPL signals. The backscattered TEPL signals from a sample were collected by the same objective lens. Note that we use high NA objective lens for efficient collection of the interlayer exciton emissions which has out-of-plane dipole moment. In addition, undesirable far-field background noise was reduced by using a pinhole in the detection scheme. TEPL signals (633 nm cut-off) were then sent to a spectrometer (f = 320 mm, 150 g/mm, ~1.6 nm spectral resolution, Monora320i, Dongwoo Optron) and finally imaged onto a thermoelectrically cooled charge-coupled device (CCD, DU971-BV, Andor) to obtain TEPL spectra. For hyperspectral nano-imaging, TEPL spectra at each pixel were recorded during an AFM scanning by a digital controller (Solver next SPM controller, NT-MDT) based on the Au tip attached on a quartz tuning fork. The Au tip (apex radius of ~10 nm) was prepared by the refined electrochemical etching protocol [43] and attached to a tuning fork with a super glue. The tip-sample distance was regulated by the shear-force feedback through monitoring the changing dithering amplitude of the tuning fork/tip assembly.

**Tip-induced pressure-engineering in heterobilayer**
To perform the nanoscale pressure-engineering of the heterobilayer using the Au tip, we gradually changed the setpoint of the shear-force feedback. To modify the electronic bandstructure (Fig. 3), we gradually lowered the setpoint to ~ 75 % of the initial oscillating amplitude to induce ~GPa pressure to the crystal structures.

**Simulations of electronic bandstructures in TMD hetero- and homo-bilayers**
To calculate electronic band structures, we performed plane-wave density functional theory (DFT) employing projector augmented-wave (PAW) potentials [44, 45] implemented in the in the VASP package [46, 47]. The electronic exchange-correlation interactions were treated using generalized gradient approximation (GGA) with the method of Perdew-Burke-Ernzerhof (PBE) [48]. In the first step, we performed a full geometry optimization for the bulk structure of the heterostructure with a plane-wave cutoff energy of 400 eV and a 10 × 10 × 10 grid of Monkhorst-Pack points until the change of the total energy between two relaxation steps was smaller than 10$^{-4}$ eV. Then we used the optimized geometry to construct the unit cell of the bilayer heterostructure. A distance of 20 Å was considered between the bilayers to avoid any interaction between them. To simulate the hydrostatic pressure, we added a small change to the interlayer distance followed by a self-consistent field calculation to find the electronic ground state with a plane-wave cutoff energy of 400 eV on a 15 × 15 × 1 k-point grid. The tolerance of self-consistent field calculation was set to $10^{-5}$ eV for both the total energy change and the band-structure-energy change between two steps. Spin-orbit coupling was not included in our calculation.

**Reference**

[1] Yu, W. J., Liu, Y., Zhou, H., Yin, A., Li, Z., Huang, Y., & Duan, X. (2013). Highly efficient gate-tunable photocurrent generation in vertical heterostructures of layered materials. *Nature nanotechnology*, *8*(12), 952-958.

[2] Britnell, L., Ribeiro, R. M., Eckmann, A., Jalil, R., Belle, B. D., Mishchenko, A., ... & Novoselov, K. S. (2013). Strong light-matter interactions in heterostructures of atomically thin films. *Science*, *340*(6138), 1311-1314.

[3] Ciarrocchi, A., Tagarelli, F., Avsar, A., & Kis, A. (2022). Excitonic devices with van der Waals heterostructures: valleytronics meets twistronics. *Nature Reviews Materials*, 1-16.

[4] Kennes, D. M., Claassen, M., Xian, L., Georges, A., Millis, A. J., Hone, J., ... & Rubio, A. (2021). Moiré heterostructures as a condensed-matter quantum simulator. *Nature Physics*, *17*(2), 155-163.

[5] Policht, V. R., Russo, M., Liu, F., Trovatello, C., Maiuri, M., Bai, Y., ... & Cerullo, G. (2021). Dissecting Interlayer Hole and Electron Transfer in Transition Metal Dichalcogenide Heterostructures via Two-Dimensional Electronic Spectroscopy. *Nano Letters*, *21*(11), 4738-4743.

[6] Rivera, P., Schaibley, J. R., Jones, A. M., Ross, J. S., Wu, S., Aivazian, G., ... & Xu, X. (2015). Observation of long-lived interlayer excitons in monolayer $MoSe_2$–$WSe_2$ heterostructures. *Nature communications*, *6*(1), 1-6.

[7] Nayak, P. K., Horbatenko, Y., Ahn, S., Kim, G., Lee, J. U., Ma, K. Y., ... & Shin, H. S. (2017). Probing evolution of twist-angle-dependent interlayer excitons in $MoSe_2$/$WSe_2$ van der Waals heterostructures. *ACS nano*, *11*(4), 4041-4050.

[8] Rivera, P., Yu, H., Seyler, K. L., Wilson, N. P., Yao, W., & Xu, X. (2018). Interlayer valley excitons in heterobilayers of transition metal dichalcogenides. *Nature nanotechnology*, *13*(11), 1004-1015.

[9] Yu, H., Liu, G. B., Tang, J., Xu, X., & Yao, W. (2017). Moiré excitons: From programmable quantum emitter arrays to spin-orbit–coupled artificial lattices. *Science advances*, *3*(11), e1701696.

[10] Unuchek, D., Ciarrocchi, A., Avsar, A., Watanabe, K., Taniguchi, T., & Kis, A. (2018). Room-temperature electrical control of exciton flux in a van der Waals heterostructure. *Nature*, *560*(7718), 340-344.

[11] Jauregui, L. A., Joe, A. Y., Pistunova, K., Wild, D. S., High, A. A., Zhou, Y., ... & Kim, P. (2019). Electrical control of interlayer exciton dynamics in atomically thin heterostructures. *Science*, *366*(6467), 870-875.

[12] Wang, Z., Rhodes, D. A., Watanabe, K., Taniguchi, T., Hone, J. C., Shan, J., & Mak, K. F. (2019). Evidence of high-temperature exciton condensation in two-dimensional atomic double layers. *Nature*, *574*(7776), 76-80.

[13] Zhang, S., Li, B., Chen, X., Ruta, F. L., Shao, Y., Sternbach, A. J., ... & Basov, D. N. (2022). Nano-spectroscopy of excitons in atomically thin transition metal dichalcogenides. *Nature Communications*, *13*(1), 1-8.

[14] Yao, K., Zhang, S., Yanev, E., McCreary, K., Chuang, H. J., Rosenberger, M. R., ... & Schuck, P. J. (2021). Nanoscale Optical Imaging of 2D Semiconductor Stacking Orders by Exciton-Enhanced Second Harmonic Generation. *arXiv preprint arXiv:2111.06955*.

[15] Siday, T., Sandner, F., Brem, S., Zizlsperger, M., Perea-Causin, R., Schiegl, F., ... & Huber, R. (2022). Ultrafast Nanoscopy of High-Density Exciton Phases in WSe2. *Nano Letters*.

[16] Tran, T. N., Kim, S., White, S. J., Nguyen, M. A. P., Xiao, L., Strauf, S., ... & Xu, Z. Q. (2021). Enhanced Emission from Interlayer Excitons Coupled to Plasmonic Gap Cavities. *Small*, *17*(45), 2103994.

[17] May, M. A., Jiang, T., Du, C., Park, K. D., Xu, X., Belyanin, A., & Raschke, M. B. (2020). Nanocavity clock spectroscopy: resolving competing exciton dynamics in WSe2/MoSe2 heterobilayers. *Nano Letters*, *21*(1), 522-528.

[18] Park, S., Kim, D., & Seo, M. K. (2021). Plasmonic photonic crystal mirror for long-lived interlayer exciton generation. *ACS Photonics*, *8*(12), 3619-3626.

[19] Kravtsov, V., Liubomirov, A. D., Cherbunin, R. V., Catanzaro, A., Genco, A., Gillard, D., ... & Iorsh, I. V. (2021). Spin–valley dynamics in alloy-based transition metal dichalcogenide heterobilayers. *2D Materials*, *8*(2), 025011.

[20] Koo, Y., Kim, Y., Choi, S. H., Lee, H., Choi, J., Lee, D. Y., ... & Park, K. D. (2021). Tip-Induced Nano-Engineering of Strain, Bandgap, and Exciton Funneling in 2D Semiconductors. *Advanced Materials*, *33*(17), 2008234.

[21] Darlington, T. P., Carmesin, C., Florian, M., Yanev, E., Ajayi, O., Ardelean, J., ... & Schuck, P. J. (2020). Imaging strain-localized excitons in nanoscale bubbles of monolayer WSe2 at room temperature. *Nature Nanotechnology*, *15*(10), 854-860.

[22] Park, K. D., Raschke, M. B., Atkin, J. M., Lee, Y. H., & Jeong, M. S. (2017). Probing Bilayer Grain Boundaries in Large-Area Graphene with Tip-Enhanced Raman Spectroscopy. *Advanced Materials*, *29*(7), 1603601.

[23] Park, K. D., Jiang, T., Clark, G., Xu, X., & Raschke, M. B. (2018). Radiative control of dark excitons at room temperature by nano-optical antenna-tip Purcell effect. *Nature nanotechnology*, *13*(1), 59-64.

[24] Sichert, J. A., Tong, Y., Mutz, N., Vollmer, M., Fischer, S., Milowska, K. Z., ... & Feldmann, J. (2015). Quantum size effect in organometal halide perovskite nanoplatelets. *Nano letters*, *15*(10), 6521-6527.

[25] Ji, J., Delehey, C. M., Houpt, D. N., Heighway, M. K., Lee, T., & Choi, J. H. (2020). Selective Chemical Modulation of Interlayer Excitons in Atomically Thin Heterostructures. *Nano letters*, *20*(4), 2500-2506.

[26] Zhang, Y., Voronine, D. V., Qiu, S., Sinyukov, A. M., Hamilton, M., Liege, Z., ... & Scully, M. O. (2016). Improving resolution in quantum subnanometre-gap tip-enhanced Raman nanoimaging. *Scientific reports*, *6*(1), 1-9.

[27] Shanks, D. N., Mahdikhanysarvejahany, F., Muccianti, C., Alfrey, A., Koehler, M. R., Mandrus, D. G., ... & Schaibley, J. R. (2021). Nanoscale trapping of interlayer excitons in a 2D semiconductor heterostructure. *Nano letters*, *21*(13), 5641-5647.

[28] Zhao, W., Regan, E. C., Wang, D., Jin, C., Hsieh, S., Wang, Z., ... & Wang, F. (2021). Dynamic Tuning of Moiré Excitons in a WSe2/WS2 Heterostructure via Mechanical Deformation. *Nano Letters*, *21*(20), 8910-8916.

[29] Ciarrocchi, A., Unuchek, D., Avsar, A., Watanabe, K., Taniguchi, T., & Kis, A. (2019). Polarization switching and electrical control of interlayer excitons in two-dimensional van der Waals heterostructures. *Nature photonics*, *13*(2), 131-136.

[30] Ma, X., Fu, S., Ding, J., Liu, M., Bian, A., Hong, F., ... & He, D. (2021). Robust Interlayer Exciton in WS2/MoSe2 van der Waals Heterostructure under High Pressure. *Nano Letters*, *21*(19), 8035-8042.

[31] Xia, J., Yan, J., Wang, Z., He, Y., Gong, Y., Chen, W., ... & Shen, Z. (2021). Strong coupling and pressure engineering in WSe2–MoSe2 heterobilayers. *Nature Physics*, *17*(1), 92-98.

[32] Lee, H., Woo, J. Y., Park, D. Y., Jo, I., Park, J., Lee, Y., ... & Park, K. D. (2021). Tip-Induced Strain Engineering of a Single Metal Halide Perovskite Quantum Dot. *ACS nano*, *15*(5), 9057-9064.

[33] Benimetskiy, F. A., Sharov, V. A., Alekseev, P. A., Kravtsov, V., Agapev, K. B., Sinev, I. S., ... & Iorsh, I. V. (2019). Measurement of local optomechanical properties of a direct bandgap 2D semiconductor. *APL Materials*, *7*(10), 101126.

[34] Currie, M., Hanbicki, A. T., Kioseoglou, G., & Jonker, B. T. (2015). Optical control of charged exciton states in tungsten disulfide. *Applied Physics Letters*, *106*(20), 201907.

[35] Lundt, N., Cherotchenko, E., Iff, O., Fan, X., Shen, Y., Bigenwald, P., ... & Schneider, C. (2018). The interplay between excitons and trions in a monolayer of MoSe2. *Applied Physics Letters*, *112*(3), 031107.

[36] Ghods, S., & Esfandiar, A. (2021). Plasmonic enhancement of photocurrent generation in two-dimensional heterostructure of WSe2/MoS2. *Nanotechnology*, *32*(32), 325203.

[37] Deilmann, T., & Thygesen, K. S. (2018). Interlayer trions in the MoS2/WS2 van der Waals heterostructure. *Nano letters*, *18*(2), 1460-1465.

[38] Choi, C., Huang, J., Cheng, H. C., Kim, H., Vinod, A. K., Bae, S. H., ... & Wong, C. W. (2018). Enhanced interlayer neutral excitons and trions in trilayer van der Waals heterostructures. *npj 2D Materials and Applications*, *2*(1), 1-9.

[39] Zhang, H., Wei, J., Zhang, X. G., Zhang, Y. J., Radjenovica, P. M., Wu, D. Y., ... & Li, J. F. (2020). Plasmon-induced interfacial hot-electron transfer directly probed by Raman spectroscopy. *Chem*, *6*(3), 689-702.

[40] Reddy, H., Wang, K., Kudyshev, Z., Zhu, L., Yan, S., Vezzoli, A., ... & Meyhofer, E. (2020). Determining plasmonic hot-carrier energy distributions via single-molecule transport measurements. *Science*, *369*(6502), 423-426.

[41] Tagliabue, G., DuChene, J. S., Abdellah, M., Habib, A., Gosztola, D. J., Hattori, Y., ... & Atwater, H. A. (2020). Ultrafast hot-hole injection modifies hot-electron dynamics in Au/p-GaN heterostructures. *Nature Materials*, *19*(12), 1312-1318.

[42] Jain, A., Bharadwaj, P., Heeg, S., Parzefall, M., Taniguchi, T., Watanabe, K., & Novotny, L. (2018). Minimizing residues and strain in 2D materials transferred from PDMS. *Nanotechnology*, *29*(26), 265203.

[43] Neacsu, C. C., Steudle, G. A., & Raschke, M. B. (2005). Plasmonic light scattering from nanoscopic metal tips. *Applied Physics B*, *80*(3), 295-300.

[44] Blöchl, P. E. (1994). Projector augmented-wave method. *Physical review B*, *50*(24), 17953.

[45] Kresse, G., & Joubert, D. (1999). From ultrasoft pseudopotentials to the projector augmented-wave method. *Physical review b*, *59*(3), 1758.

[46] Kresse, G., & Furthmüller, J. (1996). Efficient iterative schemes for ab initio total-energy calculations using a plane-wave basis set. *Physical review B*, *54*(16), 11169.

[47] Kresse, G., & Furthmüller, J. (1996). Efficiency of ab-initio total energy calculations for metals and semiconductors using a plane-wave basis set. *Computational materials science*, *6*(1), 15-50.

[48] Perdew, J. P., Burke, K., & Ernzerhof, M. (1996). Generalized gradient approximation made simple. *Physical review letters*, *77*(18), 3865.

[49] Kravtsov, V., Berweger, S., Atkin, J. M., & Raschke, M. B. (2014). Control of plasmon emission and dynamics at the transition from classical to quantum coupling. *Nano letters*, *14*(9), 5270-5275.